\RequirePackage{fix-cm}
\documentclass[twocolumn]{svjour3}
\smartqed
\usepackage{graphicx}
\usepackage{color}

\begin{document}

\title{A biaxial apparatus for the study of heterogeneous and intermittent strains in granular materials}

\author{Antoine Le Bouil         \and
        Axelle Amon \and
        Jean-Christophe Sangleboeuf \and
        Herv\'e Orain \and
        Pierre B\'esuelle \and
        Gioacchino Viggiani \and
        Patrick Chasle \and
        J\'er\^ome Crassous
}

\institute{A. Le Bouil \and A. Amon \and P. Chasle \and J. Crassous   \at
              IPR - UMR 6251, University Rennes 1,
              F-35042 Rennes Cedex, France\\
                         \email{axelle.amon@univ-rennes1.fr} \\
                         \email{jerome.crassous@univ-rennes1.fr} \\
                                    \and
           J.-C. Sangleboeuf \and
           H.Orain \at
              LARMAUR - ERL CNRS 6274, University Rennes 1,
              F-35042 Rennes Cedex, France\\
           \and
            P. B\'esuelle \and
            G. Viggiani \at
                LUJF-Grenoble 1 / Grenoble-INP / CNRS UMR 5521, Laboratoire 3SR, Grenoble, France\\
}
\date{Received: date / Accepted: date}
\maketitle

\begin{abstract}
We present an experimental apparatus specifically designed to investigate the
precursors of failure in granular materials. A sample of granular
material is placed between a latex membrane and a glass plate. A  confining effective pressure is applied by applying vacuum to the sample. Displacement-controlled compression is applied in the vertical direction, while the specimen deforms in plane strain. A Diffusing Wave Spectroscopy visualization setup gives
access to the measurement of deformations near the glass plate. After
describing the different parts of this experimental setup, we present
a demonstration experiment where extremely small (of order $10^{-5}$) heterogeneous strains are measured during the loading process.

\end{abstract}

\section{Introduction}
\label{intro}

\begin{figure*}[htbp]
\begin{center}
\includegraphics[width=0.9\textwidth]{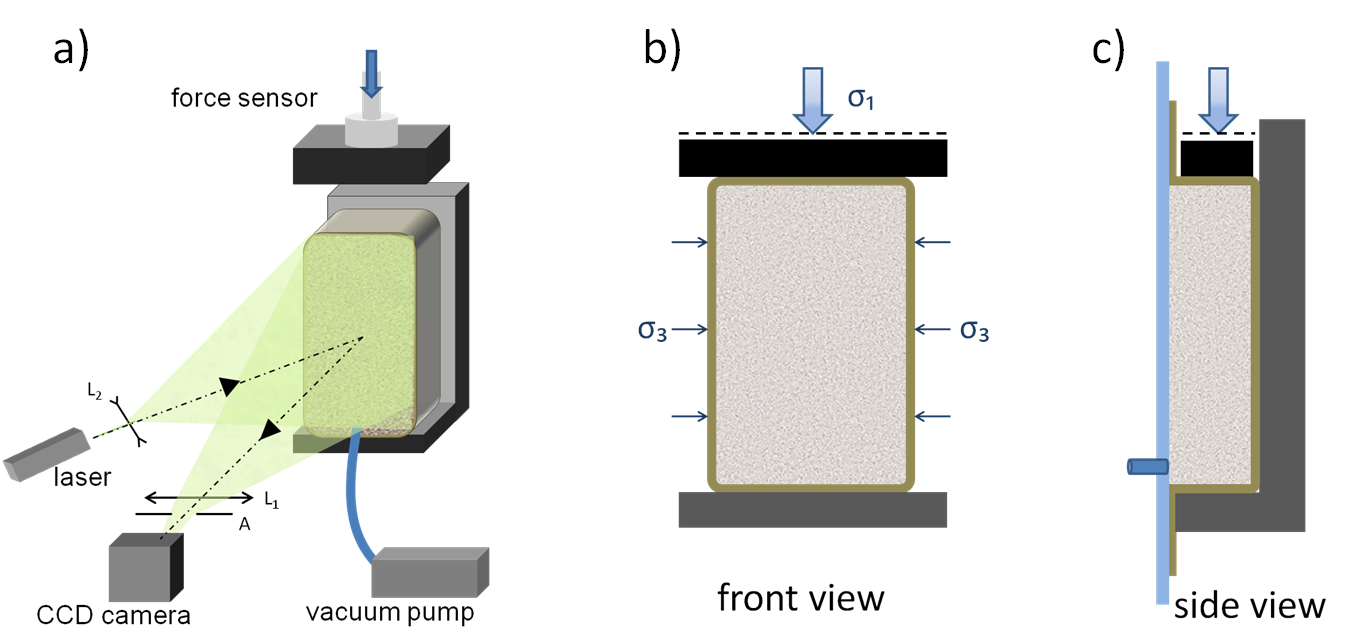}
\end{center}
\caption{Sketch of the experimental setup. The material is placed
  between a glass plate and a latex membrane. (a) The front is
  illuminated by a laser beam which is expended by the lens $L_{2}$ in
  order to light the whole sample. Multiple light scattering occurs in
  the grains assembly and part of the light is backscattered. The
  front side of the sample is imaged by the lens $L_{1}$ on the CCD
  camera. The size of the speckle spots is controlled by the diaphragm
  aperture A. The vacuum pump maintains a depressurization inside the
  sample at the origin of the stress $\sigma_3$ on the lateral
  walls. A force sensor on the top of the loading plate measures the
  stress $\sigma_1$ corresponding to the imposed displacement. (b)
  Front view with the applied stresses. (c) Side view showing the
  plane strain configuration: the displacements are blocked by the
  front (glass plate) and back walls.}
\label{fig1}
\end{figure*}

Describing the mechanical behavior of granular materials is a very challenging task, especially when dealing with strain localization phenomena that eventually leads to macroscopic failure. Strain localization (often referred to as shear banding) has quite a practical relevance, as stability and deformation characteristics of earth structures are often controlled by the soil behavior within the zones of localized strain~\cite{schofield,davis,nedderman}. Yet, the mechanisms responsible for the formation of these zones in granular materials are still subject to debate. While the formation of a shear band is now generally interpreted as a bifurcation problem in continuum mechanics (e.g.~\cite{rice1976,rudnicki1975}), its theoretical and numerical treatment still presents special problems for granular materials – ( see for example the monograph by Vardoulakis and Sulem ~\cite{vardoulakis} and the review paper by B\'esuelle and Rudnicki~\cite{besuelle2004}).
While the issues of orientation and thickness of shear bands have been debated for decades, other, more intricate issues have more recently attracted the interest of experimental research in this field: among the others, the occurrence of temporary or "non-persistent" modes of localization, that is, localized regions which form during the test and eventually "disappear" (see for example~\cite{hall2010}). From an experimental standpoint, temporary modes of strain localization are inherently more difficult to observe and characterize than final, persistent shear bands. Not only they transient, but they are also characterized by less intense shear strain.
Many optical methods are available for measuring strain fields in a deforming granular assembly (see~\cite{viggiani2012}). However, many of them fail to detect these temporary patterns of strain localization, because their spatial and temporal resolution is not fine enough (it should be mentioned, though, that recently the performance of (Digital Image Correlation) DIC based methods has been significantly improving thanks to the increasingly higher resolution of existing cameras). In this paper, we present a full-field method that has the capability to detect fine non persistent regions of localized strain. First we present the experimental setup, specifically designed to resolve strains as small as $10^{-5}$, as well as the interferometric method for measuring strain (Section 2). In section 3 we discuss the experimental protocols, and in Section 4 we discuss the results of a typical experiment, showing the potential of the experimental setup.

\section{Experimental setup}
\label{sec:2}
\subsection{Overview of the setup}
\label{sec:2.1}

The experimental setup consists of a plane strain apparatus coupled with a
dynamic light scattering setup. The sche\-matic drawing of the setup
is shown in fig.~\ref{fig1}. The granular material is placed between a
latex membrane and a glass plate, and a vacuum is applied in the
granular material. A confining effective pressure $\sigma_3$ results from the applied vacuum. The granular material is compressed on the
top through the displacement of a plate. The displacement of the top
plate and the force exerted on it are measured during the
compression. In front of the apparatus, on the side of the glass window,
a light scattering setup is placed. The front of the granular material
is illuminated with an extended laser beam. The light which is
scattered by the granular material is recorded by a camera imaging the
front of the sample. We analyze the scattered light in order to access
to information about the deformation of the granular material.

\subsection{Mechanical part}
\label{sec:2.2}

\begin{figure}[htbp]
\includegraphics[width=1\columnwidth]{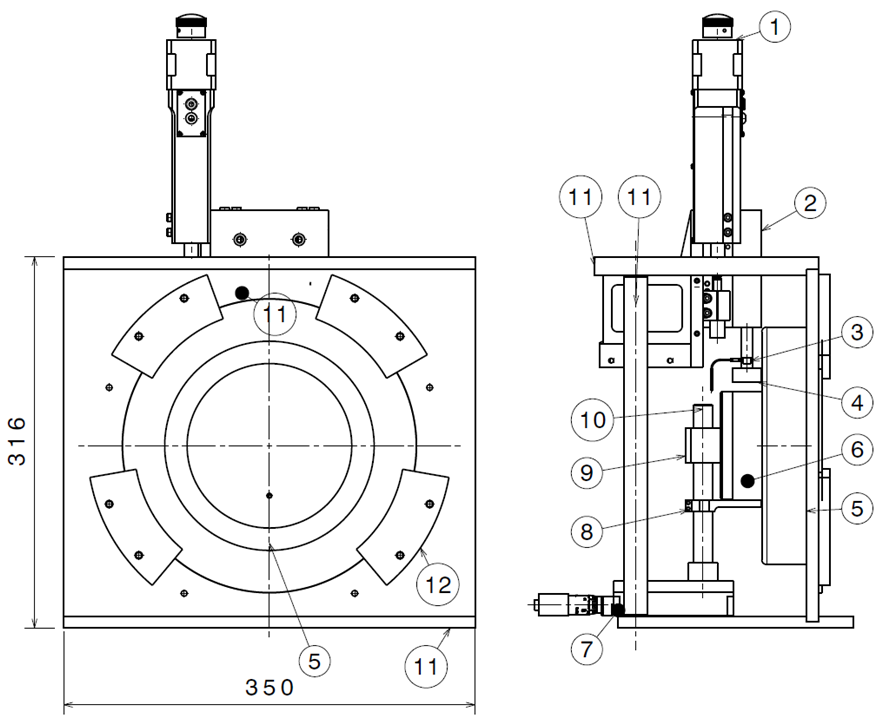}
\caption{Front (left) and side (right) views of the biaxial apparatus. \textcircled{{\tiny 1}}~motor, \textcircled{\tiny
    2}~motorized translation stage, \textcircled{\tiny 3}~force
  sensor, \textcircled{\tiny 4}~loading plate, \textcircled{\tiny
    5}~glass window, \textcircled{\tiny 6}~sample of granular
  material, \textcircled{\tiny 7}~translation stage,
  \textcircled{\tiny 8}~bottom plate, \textcircled{\tiny 9}~back
  plate, \textcircled{\tiny 10}~positioning axis, \textcircled{\tiny
    11}~structure, \textcircled{\tiny 12}~wedges holding the glass
  window.}
\label{fig2}
\end{figure}

The drawing of the mechanical part of the setup is shown on
fig.~\ref{fig2}. The granular sample (fig.~\ref{fig2}
\textcircled{\tiny 6}) of size $(85\times55\times25)$mm$^{3}$ is
submitted to a biaxial test in plane strain conditions. To limit the
gravity effects and homogenize the force along out-of-the-plane direction, the granular matter is maintained by depressurization between a
glass plate and a latex membrane. A motorized translation stage
(fig.~\ref{fig2} \textcircled{\tiny 1}-\textcircled{\tiny 2}, Thorlabs)
is used for the compression. The loading plate
(fig.~\ref{fig2}~\textcircled{\tiny 4}) is fixed on the linear
long-travel translation stage (fig.~\ref{fig2}~\textcircled{\tiny 2},
Thorlabs LNR50 Series) which is driven by a stepper motor (DC Servo
Motor Actuator, Thorlabs DRV414). The DC Motor Controller (Thorlabs
BDC101) allows to pilot the translation stage. The two main controls
are the speed of the stage (length of the step per second) and the
stop position. Between the loading plate and the translation stage a
$500$~N force sensor (fig.~\ref{fig2}~\textcircled{\tiny 3},
Measurements Specialities) gives the force applied on
the sample.

\subsection{Spatially resolved Diffusing Wave Spectroscopy}
\label{sec:2.4}

\begin{figure}[htbp]
\includegraphics[width=\columnwidth]{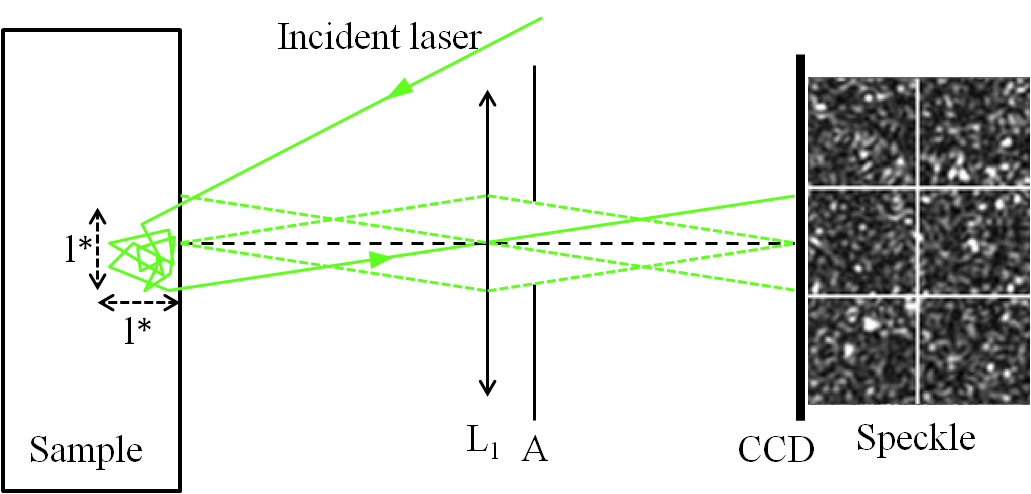}
\caption{Example of an incident ray path undergoing multiple
  scattering inside the granular media and backscattered through the
  lens and the diaphragm aperture on the camera. Right: example of
  6 meta-pixels of the speckle pattern recorded with the camera.}
\label{fig3}
\end{figure}

The measurement of the deformation of the granular material is
obtained with a spatially resolved home-made Diffusing Wave
Spectroscopy (DWS) setup described in detail
elsewhere~\cite{erpelding2008}. DWS is an interference technique using
scattering of coherent light by strongly diffusive materials. A light
beam emerging from a laser source enters the sample
(fig.~\ref{fig3}), and is scattered many times. The different exiting rays interfere on the camera sensor and
produce a speckle pattern consisting of dark and bright spots (see
right part of fig.~\ref{fig3}). A displacement of the scatterers
modifies the path length of the rays. This produces changes in the
phases of the scattered waves. Those phase variations induce
modifications of the speckle pattern. The variation of the speckle
pattern is then characterized by the computing of the correlation
function of the scattered light.

As shown in fig.~\ref{fig1}.a, we are in a backscattering geometry for
DWS, meaning that the backscattered light is collected. In practice,
the glass window side of the sample is illuminated by a laser source
of wavelength $\lambda=532$~nm and maximum output power 75~mW (Compass
215M from Coherent). The beam is expanded with the lens $L_{2}$ (see
fig.~\ref{fig1}.a) in order to illuminate the whole front face of the
sample. The image of the front is done using a lens $L_1$
of focal length 100 mm, allowing a magnification in agreement with the
whole optical system (sample-camera). The speckle pattern is recorded
with a camera (PT-41-04M60 from DALSA) with $2352\times1728$
resolution and pixel size $7.4\mu m$.

\paragraph{Speckle analysis.}
The speckle pattern is recorded by the camera. Correlations of the
scattered intensities are calculated between two images $1$ and $2$ in
the following way. First the speckle images are divided into square
regions that we called meta-pixels (see image on right of
fig.~\ref{fig3}). For each meta-pixel, the correlation function
$g^{(12)}_{I}$ between the two intensities $I_1$ and $I_2$ is
computed as,

\begin{equation}
g^{(12)}_{I} = \frac{\left\langle I_{1}I_{2}\right\rangle - \left\langle I_{1} \right\rangle\left\langle I_{2}\right\rangle}{\sqrt{\left\langle I^{2}_{1}\right\rangle - \left\langle I_{1} \right\rangle ^{2}} \sqrt{\left\langle I^{2}_{2}\right\rangle - \left\langle I_{2} \right\rangle ^{2}}},
\label{eq1}
\end{equation}
where $I_{1}$ and $I_{2}$ are the matrices of intensities of the
meta-pixel considered, and averages are performed over all the pixels
of the meta-pixel. The correlation function (eq.~(\ref{eq1})) is
normalized so that $g^{(12)}_{I}=1$ if $I_1=I_2$, and $g^{(12)}_{I}=0$ if the
intensities are uncorrelated. The link between the displacements of
the beads and the value of the correlation function may be estimated
from a model of light propagation into the granular material. As the
bead assembly strongly scatters light, rays follow random walks
inside the material. The transport mean free path, i.e. the persistent
length of the random walk of light in the material is denoted by
$l^*$. Typically, for a granular material, this optical constant is few diameters of grains. We will discuss the value of $l^*$ and its
implication on the spatial resolution of the deformation maps that we
obtain in sec.~\ref{sec:3.3}. If the deformation of the material
between the two images 1 and 2 can be considered as affine and
homogeneous on the scale of a metapixel, the correlation function
can be expressed in backscattering geometry
as~\cite{djaoui2005,crassous2007,erpelding2008}:
\begin{equation}
\left|g^{(12)}_{I}\right| \approx exp\Bigl(-\eta k l^* \sqrt{f({\bf U})}\Bigr),
\label{eq2}
\end{equation}
where $\eta$ is an optical factor of order 1, and $f({\bf U})={1 \over
  5}Tr^2({\bf U})+{2 \over 5}Tr({\bf U}^2)$ is a function of the
invariant of the strain tensor ${\bf
  U}=\frac{1}{2}\left(\frac{\partial u_{i}}{\partial
  x_j}+\frac{\partial u_{j}}{\partial x_i}\right)$.  If the bead
displacements is the sum of an affine motion with a strain tensor $\bf
U$ and an uncorrelated motion of the beads, we then
expect~\cite{bicout1994}:

\begin{equation}
\left|g^{(12)}_{I}\right| \approx exp\Bigl(-\eta k \sqrt{(l^*)^2
  f({\bf U})+\langle \Delta r^2 \rangle}\Bigr),
\label{eq2bis}
\end{equation}
with $\langle \Delta r^2 \rangle$ the average of the quadratic
displacement of the uncorrelated motion of the beads.

\section{Experimental Protocol}
\label{sec:3}

Here we summarize the different steps of the protocol of fabrication
of the latex membrane and of preparation of the granular material. The
limitations and the sensitivity of both the mechanical device and the
optical setups are detailed. The control of the camera and the
acquisition of images, the force sensor monitoring and the
displacement control are all ensured by a Labview interface on a
computer.

\subsection{Preparation of the membrane and granular sample}
\label{sec:3.1}

\begin{figure}[htbp]
\includegraphics[width=\columnwidth]{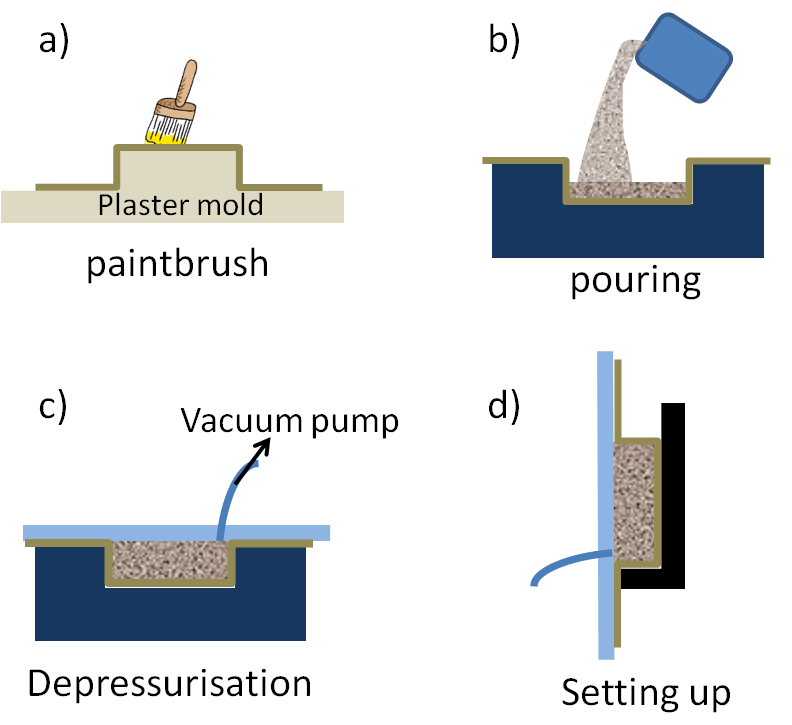}
\caption{Steps for preparing a sample. (a) Confection of the latex
  membrane on a plaster mold. (b) The latex membrane is fitted into a
  counter mold and the granular material is poured and possibly
  compacted. (c) The glass plate is laid overall, and partial vacuum
  is created in the sample. (d) The sample is placed on the biaxial
  apparatus.}
\label{fig4}
\end{figure}

As shown in fig.~\ref{fig4}.a, the latex membrane is made by spreading
successive layers of liquid latex with a paintbrush on a plaster mold
of the required dimensions. After drying it generates a planar
membrane with a hollow of the size of the sample. The same membrane
can be used for several tests. To prepare the sample, the membrane is
laid in a counter-mold horizontally and the granular material is poured
in the hollow (fig.~\ref{fig4}.b). A compaction protocol is applied to
the pile according to the desired volume fraction. The glass plate is
then placed over the membrane (fig.~\ref{fig4}.c). A flexible pipe
connects the vacuum pump to the glass plate and the sample is pressed
on the glass instantly by depressurization. We use vacuum grease between
the annular part of the latex membrane and the glass plate for better
sealing and adherence. The grease is viscous and does not creep into the bulk solid during the whole time of the experiment. A pressure sensor and an air valve allow the
measurement and control of the depressurization using a feedback
loop. Then the window (fig.~\ref{fig2}~\textcircled{\tiny 5}), bearing
the sample, is fixed on the structure of the biaxial apparatus using
wedges (fig.~\ref{fig2}~\textcircled{\tiny 12}). Because of residual
air leakages and long duration experiments, the pump and the pressure
control system are active during the whole loading experiment. To
achieve plane strain conditions, the back plate and the bottom plate
are brought in contact with the sample using a translation stage
(fig.~\ref{fig2}~\textcircled{\tiny 7}). The loading is applied using
a plate larger than the sample in the $\sigma_3$ direction (more than
55 mm, see fig.~\ref{fig1}.b), and slightly smaller than the sample
thickness in the other direction (less than 25 mm, see
fig.~\ref{fig1}.c). The distances between the glass
plate and the loading plate on one hand and between the back plate and
the loading plate on the other hand are approximately of 1~mm. Our
apparatus does not give the position of the granular material with
respect to the motorized translation stage \textcircled{\tiny 2}.

\subsection{Mechanical stiffness of the system}
\label{sec:3.2}

Several tests have been conducted to characterize the apparatus. The
range of the force sensor is limited to 500~N and its calibration has
been made using known loads. Its stiffness has been investigated and
the spring constant of the sensor has been measured in two different
ways. First, using a stiff micrometer stage we have imposed a known
deformation to the sensor, and have measured the force on the
sensor. We obtained a stiffness for the sensor of
$k_{sens}=(4.3\pm0.2) \times 10^{6}$~N.m$^{-1}$. We have also tracked
the deformation of the sensor by following the displacement of markers
above and below the sensor using a method described
in~\cite{chean2011}, and have measured $k_{sens}=(5\pm1) \times
10^{6}$~N.m$^{-1}$. Other tests were conducted to measure the
stiffness of the bottom plate and the stiffness of the whole loading
device. The stiffness of the whole system has been tested in a Hertz
contact configuration and is found to be $k_{sys}=7.3 \times
10^{5}$~N.m$^{-1}$. The stiffness of a block of material of section
$S$, length $L$ and of Young modulus $E$ is $k_{mat}=S E /L$. For $S
\approx 1.4 \times 10^{-3}$~m$^2$ and $L \approx 85$~mm,
$k_{sys}=k_{mat}$ for $E=44$~MPa. This gives an estimation of the
limit value of the Young modulus above which the correction due to the
apparatus stiffness must be taken into account. The motorized
linearized stage is limited to one step of 1~$\mu$m per second and to
a maximal loading of $250~N$. Tests confirm this value, with an
activation of security around $240~N$. Estimation of the membrane
rigidity from the values of latex elastic modulus show that the
elasticity of the membrane is negligible.

\subsection{Optical protocol and sensitivity}
\label{sec:3.3}

The first optical issue is to acquire a suitable speckle pattern. The
focal length of the lens and the distances between the sample, the
lens and the camera have to be in agreement with the magnification
allowing to image the whole surface of the sample on the camera. Then the diaphragm aperture (see fig.~\ref{fig3}) allows adjustment of the size of the speckle spots. This size is determined from the
Fourier transform of the image of intensity. The size of the speckle
spot is fixed to $\approx 2.3$~pixels which is a good compromise
between the necessity to have the maximum number of speckle spots, and
the need for the speckle spots to be at least few pixels.

Concerning the time resolution, the stepper motor frequency is limited
to one step per second, the acquisition rate of the images to $4$
frames per second, and we acquire $1$ image per motor step. The
correlation maps are calculated between 2 images using
eq.~(\ref{eq1}). These two images can be spaced out by a specific time
interval.

The spatial resolution of the deformation map that we obtain is
dependent on both the camera used and of the granular material. As for
the camera, the correlation functions must be calculated on large
enough areas in order to minimize the statistical error when
evaluating eq.~(\ref{eq1}). In practice, the noise on the correlation
function is kept to an acceptable level if the size of a meta-pixel is
at least $16 \times 16$~pixels. With a speckle size of $2.3$~pixels,
this corresponds to $\sim 50$ speckle spots per metapixel. Concerning
the material, the resolution is limited by the light scattering
process inside the granular media. Photons that enter the granular
material at a given point typically exit the media at a distance
smaller or of order $l^*$ and have mainly probed zones inside the
material of size $l^*$. So, as shown on fig.~\ref{fig3}, a photon that
emerges from a given point of the material have explored a volume of
order $(l^{*})^{3}$. The information about the deformation of the
material is thus naturally averaged over this volume as shown in
fig.~\ref{fig3}. The spatial resolution is thus limited to $l^*$ by
the scattering process. The optimal choice of the optical setup is
when a meta-pixel on the camera is the image of an area of order $l^*
\times l^*$ on the object. If we call $L$ the largest dimension of the
material sample, the optimum of resolution is attained with our $2352$
pixel width camera when $L/l^* = 2352/16 \simeq 150$. For a sample
size of $L=85$~mm, this corresponds to an optimal value of $l^* \simeq
0.6$~mm. For spherical glass beads, model of light propagation into
granular material~\cite{crassous2007} and
experiments~\cite{crassous2007,djaoui2005,amon2013} show that $l^*
\simeq 3.3$~$d$, with $d$ the mean bead diameter. For Fontainebleau
sand, we measured $l^* \simeq 2.0~d$~\cite{amon2013}. So optimal value
for grains diameter is typically $d=200-300\mu m$ depending on the
material.

It should be stressed that those resolution considerations hold for
the best spatial resolution. With this resolution, one meta-pixel
corresponds to deformation that are averaged on typical sizes of $l^*$
i.e. of few $d$. Larger beads may be also used: we made DWS studies
with glass beads of diameter $d=500~\mu$m in a different experimental
geometry~\cite{crassous2008,richard2008}. Smaller beads may also be
used but with a spatial resolution larger than $l^*$. Keeping the same
spatial resolution may however be achieved by zooming on a smaller
part of the sample, or with an image acquisition device with better
spatial resolution such as high resolution camera.

\section{A demonstration experiment}
\label{sec:4}

\begin{figure*}
\includegraphics*[width=1.\textwidth]{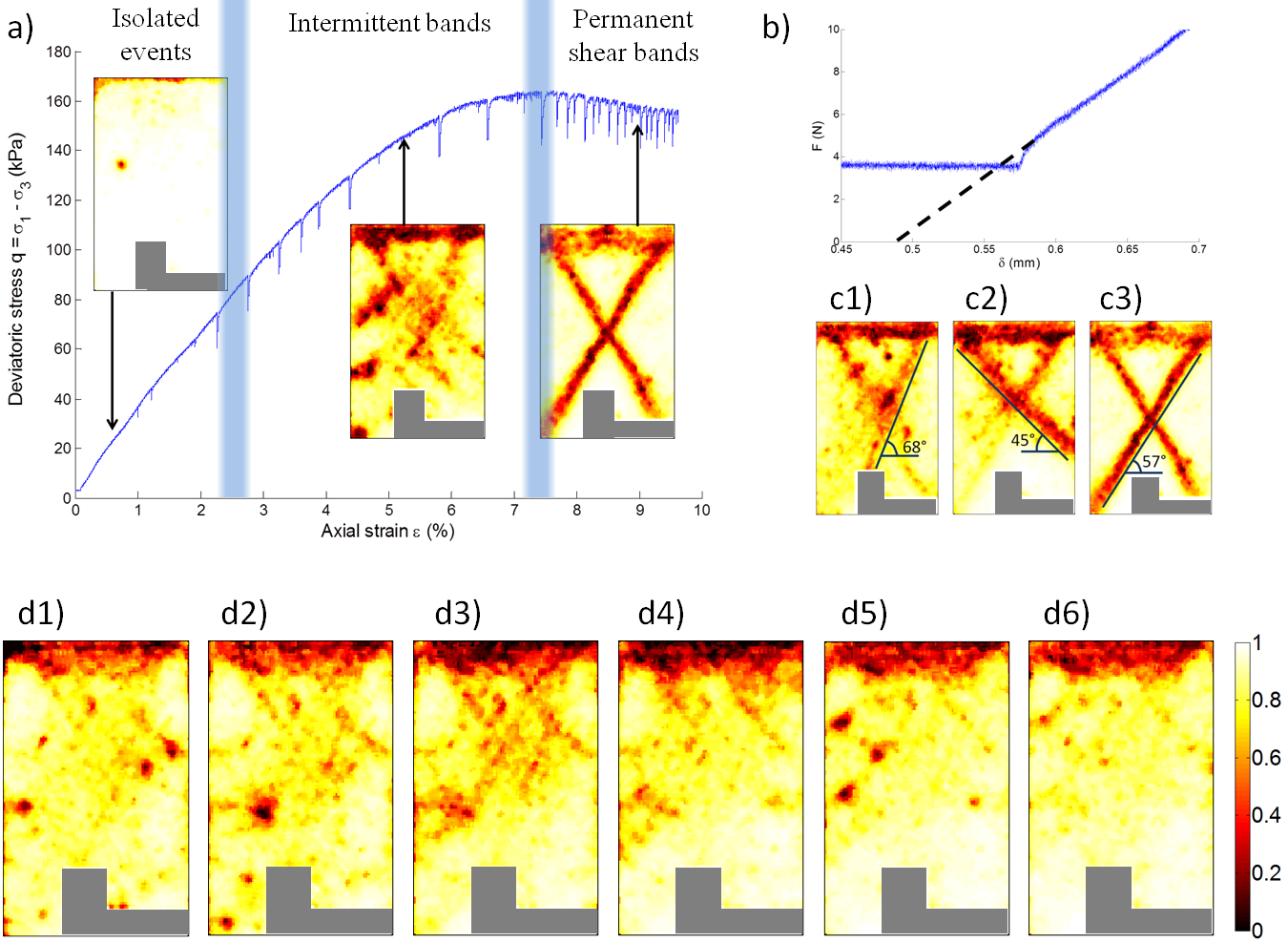}
\caption{(a) Loading curve: Difference of stresses
  $q=\sigma_{1}-\sigma_{3}$ versus vertical strain $\epsilon$ for a
  confining stress $\sigma_3=30$~kPa and associated correlation maps
  computed for an incremental vertical strain of $\Delta \epsilon
  \simeq1.2\cdot10^{-5}$. (b) Zoom of the beginning of the loading
  process. The offset position for the displacement is defined from
  the extrapolation at zero applied force. (c) Different maps of
  correlation showing deformation bands with different angles of
  inclination (see text). (d) Successive snapshots of deformation
  showing the intermittency of the deformation map (see text) and the correlation function $g_{I}$ colorscale.}
\label{fig5}
\end{figure*}

In order to show the potential of the apparatus that we built, we
propose here a demonstration experiment where we compress a model
granular system made of cohesionless nearly spherical glass beads.

\subsection{Settings}
\label{sec:4.1}
We used for this experiment glass beads of $200-300 \mu m$ diameter in
a membrane of dimension $85$~mm height, $55$~mm width and $25$~mm depth. Concerning the
preparation of the sample, the grains are poured inside the membrane
and then softly shaken by hand for a regular compaction before the
depressurization. The confining pressure is fixed to a value
$\sigma_3=30$~kPa. Using the mean value of the grains diameter, the
transport mean free path can be estimated to $l^{*}\approx
0.66~mm$. The obtained image is divided in squares of $16\times16$
pixels which correspond to squares of $(0.64\times0.64) mm^{2}$ on the
sample. The typical noise of the correlation function $g_{I}$ is
$0.03$. For a quasi-static loading we chose for the velocity of the
top plate $1 \mu m/s$ which is approximately an vertical strain of
$1.2\cdot10^{-5}$ between $2$ motors steps. We acquire one frame per
second.

\subsection{Stress strain curve}
\label{sec:4.2}
The loading curve showing the difference of stresses versus the vertical
strain is presented on fig.~\ref{fig5}.a. The vertical strain is obtained
as $\epsilon=(\delta-\delta_0)/L$, with $L=85$~mm and $\delta$ the
loading plate displacement. The value of the offset of displacement
$\delta_0$ for which the vertical deformation $\epsilon$ is null is
determined by the linear extrapolation of the force-displacement curve
at zero force (see fig.\ref{fig5}.b). The difference of stresses $q=\sigma_1-\sigma_3$ is obtained using $q=F/S$, with $F$ the measured
force, and $S=1.37 \times 10^{-3}~m^2$ the section of the
material. The stress versus strain curve of the material is typical of
soil mechanics experiments, with first a quasi-linear part. Then the slope decreases
until a maximum of the curve is reached, followed by a slow stress
relaxation. We also observe small stress drops during the loading
process. Such stress drops are not systematically observed. In some similar experiments not shown here, we do not observe them during the loading. Such drops are reported in the literature~\cite{amon2012}, they correlate typically with an instability in part of the force network.
This is in agreement with the fact that we do not observe them during the compression of a
bulk elastic material of same Young modulus value.

\subsection{Strain maps}
\label{sec:4.3}
Some typical maps of correlations between two successive images
(corresponding to an vertical strain increment of $\Delta \epsilon \simeq
1.2\cdot10^{-5}$) are shown on fig.~\ref{fig5}.d.

These are roughly representative of the three distinct ways to deform. At the beginning of
the loading, we first observe a background of deformation with some
isolated reorganizations already observed and described in another
experiment~\cite{amon2012}. At a vertical deformation of order $\epsilon
\simeq 0.02$, we begin to see some decorrelated areas that occur in
the form of small bands. Such bands are intermittent: they appear and
disappear as the loading proceeds. The positions of such bands seem to
be arbitrary: the ending points may be on the boundaries of the sample
or not. Those bands may cross each other, or a band may start from
another one. We observe several orientations (see fig.~\ref{fig5}.c1
and c2) from $\simeq 45^\circ$ to $\simeq 68^\circ$. A $45^\circ$
angle corresponds to the Roscoe angle $\theta_R$ for non dilating
material. Those bands share some analogies with the micro-bands
described by Kuhn~\cite{kuhn1999} or Tordesillas~\cite{tordesillas2008} in numerical Discrete Element Method (DEM) simulations. More
inclined bands with a angle $\simeq 68^\circ$ are also observed. This
angle is in agreement with the angle of shear bands predicted by the
Mohr-Coulomb criterium $\theta_C=45^\circ + \phi/2 \simeq 68^\circ$,
with $\sin \phi=(\sigma_1-\sigma_3)/(\sigma_1+\sigma_3) \simeq 0.733$
at the maximum load.  At the end of the loading, we observe two
permanent symmetric bands which start from the corners of the sample,
and form the shape of the letter X. Their inclination is $\simeq
57^\circ$ (see fig.\ref{fig5}.c3). We note $\theta_G$ this geometric
angle. Such symmetric bands are widely reported in the
literature. Near the loading peak the situation is confused,
displaying intermittent bands at $\theta_R$ and $\theta_C$ but also
some parts of the future permanent bands at $\theta_G$.

\subsection{Intermittency}
\label{sec:4.4}
In conventional biaxial test experiments, the map of strain is
measured during axial (vertical) strain increments of typically $\Delta \epsilon
\sim 10^{-3}$. The great sensitivity of our optical method for strains
measurement allows the determination of strain for vertical deformation
increments as small as typically $\Delta \epsilon \sim 10^{-5}$. As a
consequence we are able to investigate the instantaneous
micro-deformations of the sample. This is an important difference with
more conventional setup. The instantaneous deformation seems to show
important temporal fluctuations during the loading. Figure~\ref{fig5}.d shows six consecutive instantaneous maps of
deformation around the mean vertical strain $\epsilon_0=3.74\cdot
10^{-2}$. The deformation maps correspond to successive vertical strain
increment, i.e. the image $d_i$, for $i=1$ to $6$ corresponds to map
of deformation between vertical strain $\epsilon_0+(i-1)\Delta \epsilon$
and $\epsilon_0+i\Delta \epsilon$ with $\Delta \epsilon \simeq
1.2\cdot10^{-5}$. As it can be seen in fig.~\ref{fig5}.d the areas of
instantaneous deformation fluctuate strongly. The typical size of the
heterogeneities and the inclination of the bands of localized strain are
roughly the same during the snapshot sequence. However the positions
of these zones fluctuate strongly from one image to the
other. This seems to indicate that this part of the loading is
associated to some kind of intermittent deformation of the granular
material. In contrast, the shear bands observed after the load peak
are rather permanent, and do not show very important fluctuations.

\section{Conclusions and future work}
\label{sec:5}
We presented in this paper a biaxial apparatus that designed to measure tiny deformations in granular samples. By analysing the
evolution of the scattered light, we are able to obtain a map of the
deformation of the granular material. The level of correlation is
related to the local affine deformation of the material and to the
fluctuations of positions around this affine deformation. Finally a
test experiment on a granular material composed of glass spheres shows
that deformation modes during an increment of vertical deformation are
very various.

The obtained information about the deformations of the material is different from similar experiments using more conventional DIC methods see
e.g.~\cite{desrues2004,rechenmacher2006}). We do not track
individually the positions of the particles. Moreover, information
on the separate component of the strain tensor are not obtained: we are not,
for example, able to distinguish between small dilatation or shear of
the material. However, the ability to measure small incremental
deformations allows to obtain a very different information on the
material deformation. The emergence of localization may be tracked at
the early stages of the deformation where conventional DIC is limited
to deformation during large strain increment near the load peak, though evolutions are in progress with very high resolution cameras. If
necessary, the tracking of individual particles on a part of the front
face may be obtained to complement the DWS interferometric
imaging. For this, images in white light have to be recorded by the
same or by another camera.

We focused in this paper on the description of the apparatus and on
the experimental methods. Our demonstration experiment shows a very
complex road leading to the failure, with numerous regimes of shear
bands inclinations. In future work, we will investigate experimentally
the evolution of those intermittent micro-failures toward the final
rupture in a simple nearly mono-disperse and non cohesive granular
material made of glass beads. Studies on cohesive material or on the
nucleation of failure from a defect inside the material may also be
considered with our apparatus.

\begin{acknowledgements}
We acknowledge the financial supports of ANR "STABINGRAM"
No. 2010-BLAN-0927-01 and R\'egion Bretagne "MIDEMADE". We thank
Jean-Charles Potier for mechanical design, Eric Robin for help with
the measurement of apparatus stiffness, and Sean McNamara for fruitful
discussions.
\end{acknowledgements}

\bibliography{Granmat_2013}{}
\bibliographystyle{unsrt}
\end{document}